# Two superconducting components with different symmetries in Nd$_{1-x}$Sr$_x$NiO$_2$ films

Qiangqiang Gu[1], Yueying Li[2], Siyuan Wan[1], Huazhou Li[1], Wei Guo[2], Huan Yang[1], Qing Li[1], Xiyu Zhu[1], Xiaoqing Pan[3], Yuefeng Nie[2]* & Hai-Hu Wen[1]*

[1]National Laboratory of Solid State Microstructures and Department of Physics, Center for Superconducting Physics and Materials, Collaborative Innovation Center for Advanced Microstructures, Nanjing University, Nanjing, China

[2]National Laboratory of Solid State Microstructures, Jiangsu Key Laboratory of Artificial Functional Materials, College of Engineering and Applied Sciences, Collaborative Innovation Center of Advanced Microstructures, Nanjing University, Nanjing, China

[3]Department of Materials Science and Engineering, Department of Physics and Astronomy, University of California, Irvine, CA 92697, USA

Abstract: The pairing mechanism in cuprates remains as one of the most challenging issues in the field of condensed matter physics. The unique $3d^9$ electron orbital of the Cu$^{2+}$ ionic states in cuprates is supposed to be the major player for the occurrence of superconductivity. Recently, superconductivity at about 9 -15 K was discovered in infinite layer thin films of nickelate Nd$_{1-x}$Sr$_x$NiO$_2$ (x=0.1-0.2) which is believed to have the similar $3d^9$ orbital electrons. The key issue concerned here is about the superconducting gap function. Here we report the first set data of single particle tunneling measurements on the superconducting nickelate thin films. We find predominantly two types of tunneling spectra, one shows a V-shape feature which can be fitted very well by a *d*-wave gap function with gap maximum of about 3.9 meV, another one exhibits a full gap of about 2.35 meV. Some spectra demonstrate mixed contributions of these two components. Our results suggest that the newly found Ni-based superconductors play as close analogs to cuprates, and thus demonstrate the commonality of unconventional superconductivity.

The pairing mechanism of cuprate superconductors has been intensively studied for more than three decades, but the fundamental reason remains unresolved yet. A common feeling summarized from the past tremendous investigations tells that the $3d^9$ orbital electrons of the $Cu^{2+}$ ionic state are crucial for the formation of superconductivity. Recently superconductivity with transition temperatures of about 9~15K was observed in infinite layer nickelate thin films of $Nd_{1-x}Sr_xNiO_2$ (x=0.1-0.2) which may share the similar $3d^9$ orbital electrons as that in cuprates (*1,2*). This discovery has drawn enormous attention (*3-12*) since it may provide deeper insight about the pairing mechanism of unconventional superconductivity in cuprates. Concerning the pairing mechanism, the core issue is to know the superconducting gap function which measures the pairing interaction of the two electrons of a Cooper pair. One of the effective ways to detect the superconducting gap function is to measure the single particle tunneling spectrum in the superconducting state. In this paper, we report such investigations for the first time on the $Nd_{1-x}Sr_xNiO_2$ thin films. The results expose essential message about the superconducting pairing in this newly found superconducting system.

The $Nd_{1-x}Sr_xNiO_3$ thin films with thickness of about 6 nm are deposited with reactive molecular beam epitaxy (MBE) technique which is different from pulsed laser deposition (PLD) method used in previous work (*1-3*). Superconductivity is achieved by annealing the sample placed in an evacuated quartz tube together with a pellet of $CaH_2$. A similar procedure given in previous report (*3*) for the heat treatment is followed. The x-ray diffraction (XRD) data together with observation of superconductivity assures that the samples after post-annealing become $Nd_{1-x}Sr_xNiO_2$ (NSNO). Results reported here are obtained on samples with a nominal composition of x = 0.2. The details of synthesis and characterization of the samples will be published separately. Shown in Fig. 1A is a schematic plot of atomic structure of the NSNO. Fig.1B shows the temperature dependence of resistivity of the NSNO film measured by using a standard four-probe technique. One can see that the onset transition temperature is about 15.3 K, and zero resistivity is achieved at about 9.1 K. The rounded transition near the onset point tells that fluctuating superconductivity can occur at about 18 K. In Fig.1C,D, we show the topographic images of one film in a 2D and 3D manner, which is measured by the scanning tunneling microscope (STM). One can see that the surface is not atomically flat showing a roughness of about 1~2 nm. This large roughness may be induced by post-annealing on the films. Details about characterization of the films are given in Supplementary Information Note 1.

We then measure the tunneling spectra at different positions on the surface of the film. Due to the large surface roughness, we cannot measure spectrum with a line-scan mode in a large area, however it can be done at different locations. It is found that the tunneling spectrum is not uniform across the sample, suggesting that the film after annealing is not in a perfect epi-texture state. However, from hundreds of the measured spectra, we find that they predominantly show two types of features. One type shows a fully gapped feature with the gap of about 2.35 meV. Typical data measured at 0.35 K are shown in Fig.2A. We fit the data with the Dynes model and get a quite nice fitting, as shown by the red curve. Details about the fitting are presented in the Supplementary Information Note 2. A slight anisotropy (about 15% weight of the differential conductivity) is added to the gap function in order to have a good fit. This indicates that at least one of the bands is fully gapped. We have conducted measurements at different positions in a small area and find that the spectra all show this type behavior, the results are shown in Fig.2B. Beside the coherence peaks at about 2.35 mV, two strong side peaks show up at about 5~6 mV. The global shape of the spectrum suggests that these side peaks may correspond to some bosonic modes. In Supplementary Fig.1 we present the data measured at about 1.5 K, one can see that, due to the thermal broadening effect, the bottom of the spectrum is elevated and the coherence peaks become rounded. Another type of spectrum shows a typical V-shape feature, which is shown in Fig.3A. By doing the Dynes model fitting, as displayed by the red curve, we find that the spectrum can be nicely fitted with a *d*-wave gap. The maximum gap obtained through the fitting is about 3.9 meV. This type of spectrum can also be measured at different positions of the sample, and some time we find that this type of feature exists everywhere in a small area. In Fig.3B, we put the two types of spectra together. On the V-shape spectrum, we can see a weak kinky point at the bias voltage of about 2.35 mV which corresponds very well to the value of the spectrum with full gap. Thus we believe that the two kinds of spectra with different gap symmetries correspond to the gaps on different Fermi surfaces. The spectra with this V-shape feature measured at 1.5 K are shown in Supplementary Fig.2. Again the bottom at zero bias is elevated and the coherence peaks are smeared. In Supplementary Fig.3 we show a spectrum in a wide energy scale ($\pm$100mV). One can see that there is a clear asymmetric background beyond the gap. This feature was also observed in cuprates and was attributed to the strong correlation effect (*13*). This asymmetric can also be explained as due to the tunneling matrix problem (*14*) concerning the multiband feature in the present system. The topographic images shown in Fig.1C,D indicate strong roughness which provides the possibility for the STM tip to detect tunneling behavior along different directions at different positions. This makes actually the advantage for us to detect the superconducting gap features derived from different bands, which could be the reason for us to see two distinct gap structures at different positions. The same situation occurs in the STM measurements of $MgB_2$ bulk and film (*15*). The STM tip can detect the gap with a magnitude of about 7.1 meV on the $\sigma$-band on some grains, and can also measure the gap on the $\pi$-band with the value of 2.3 mV on other grains.

As presented by our data, two superconducting components with different gap structures have been found in Nd$_{1-x}$Sr$_x$NiO$_2$ thin films. Since the discovery of superconductivity in this system (*1*), enormous efforts have been made in order to unravel the mystery of superconductivity. Experimentally, the observation of superconductivity has been repeated by another group (*3*). Although bulk samples with the same 112 structure and proper Sr-doping have been made by some of us (*16*), but no superconductivity is observed. In addition, the bulk samples exhibit very strong insulating behavior even under a pressure up to 50.2 GPa. We argue that the absence of superconductivity may be induced by the heavy deficiency of Ni in our bulk samples. This can either strongly lower down the hole doping or act as strong scatters which localize the mobile electrons. It seems to be physically unreasonable of attributing the absence of superconductivity in bulk samples to intercalating extra hydrogen (*17*), since both the superconducting film and the bulk samples are treated with CaH$_2$ in exactly the same way.

Theoretically, electronic structures have been well calculated by several groups (*4-10*). Although the theoretical calculations about the electronic structure of NSNO differ from different groups in some way, however for the parent phase NdNiO$_2$, the major bands crossing Fermi energy are $Ni\text{-}3d_{x^2-y^2}$, $Nd\text{-}5d_{3z^2-r^2}$, and $Nd\text{-}5d_{xy}$ with some hybridizations with other bands. The $Ni\text{-}3d_{x^2-y^2}$ band contributes a large hole pocket around M point at the $k_z = 0$ cut, and evolves into an electron pocket around Z point at the $k_z = \pi$ cut due to the dispersion along $k_z$ direction. On the cutting plane at $k_z = 0$, the hole pocket looks like that of underdoped cuprate; while on the cutting plane at $k_z = \pi$, the electron pocket shrinks and becomes similar to the Fermi pocket in overdoped cuprate. Thus the $Ni\text{-}3d_{x^2-y^2}$ Fermi surface displays a von Hove singularity feature evolving from the $k_z = 0$ cut to the $k_z = \pi$ cut (*6*). We denote this $Ni\text{-}3d_{x^2-y^2}$ derived Fermi pocket as $\alpha$ pocket. The two Nd-derived orbitals

$Nd\text{-}5d_{3z^2-r^2}$ and $Nd\text{-}5d_{xy}$ contribute two three dimensional electron pockets centered at Γ (0,0,0) and A(π,π,π), which are denoted as β and γ pockets, respectively. Upon doping holes to the system, the electron-like β pocket centered around Γ point shrinks but still exists at the doping level of about 20% in NSNO, so does the γ pocket centered at A point. By considering the intra-pocket repulsive interaction within the $Ni\text{-}3d_{x^2-y^2}$ orbital, a d-wave superconducting gap is naturally expected on the α Fermi pocket (*4-6,9,10*), although certain amount of dispersion along $k_z$ exists. This pairing tendency is supposed to be the dominant one. In our experiment, a *d*-wave superconducting component with the maximum gap of about 3.9 meV is discovered. We believe this component should play the dominant role in inducing superconductivity, and combining with theoretical calculations, we intend to attribute this gap to the $Ni\text{-}3d_{x^2-y^2}$ band. In this sense, the pairing mechanism of the NSNO system may be analog to the cuprate. However, beside this similarity, we find another full gap with s-wave symmetry, which may derive from the $Nd\text{-}5d_{3z^2-r^2}$ and / or $Nd\text{-}5d_{xy}$ orbitals, namely the β and γ Fermi pockets. If we just simply follow the $d_{x^2-y^2}$ notation for the gaps in the whole momentum space, the nodal line will cut the two electron pockets, namely the β and γ pockets. This is not consistent with our observation of a full gap. Actually, there are some arguments to against the weak coupling based picture (*8,11,12,18,19*). These models postulate that correlations, multi-Hubbard interactions or doublon and holons may be crucial to determine the final pairing form function of superconductivity. We hope these models can provide explicit explanations to our observation of two superconducting components with different symmetries. In addition, one picture (*9*) concerning the inter-orbital Hubbard interaction between different bands is proposed, which expects not only a *d*-wave gap on the α pocket, but also full gaps on the β and γ ones with opposite gap signs. The latter is a bit like the s$^\pm$ pairing in many iron based superconductors. Combining our observation and this theoretical work, cartoon pictures with the Fermi surfaces and the gap structures on different cuts of $k_z$ are depicted in Fig.4. This scenario is quite interesting, and tells that not only the intra-pocket interaction, but also the inter-pocket interaction plays an important role here, leading to the orbital selective pairing. Intuitively, the pairing form in $Nd_{1-x}Sr_xNiO_2$ may serve as a bridge between the cuprate and the iron based superconductors. Because the former has only the intra-orbital interaction as the driving force for pairing, leading to the *d*-wave gap; while the latter needs the inter-orbital interaction for pairing, resulting in the orbital selective pairing and s-wave gaps with opposite signs on different Fermi pockets. At the moment, we don't know whether the gaps on the Nd-*5d* orbital derived $\beta$ and $\gamma$ pockets have gaps with opposite signs. The direct experimental evidence of the d-wave gap on the $\alpha$ pocket is also lacking. To resolve this issue, we need to do further phase-referenced quasi-particle interference experiments on single crystal samples when they are available (*20, 21*), which has been conducted successfully in iron based superconductors (*22,23*) and cuprates (*24*). Clearly, more efforts are desired in order to pin down the assignment of the superconducting gaps on different Fermi pockets.

In summary, on $Nd_{1-x}Sr_xNiO_2$ thin films, we have found predominantly two superconducting components with distinct gap symmetries. One component has a *d*-wave gap with the maximum value of about 3.9 meV, another one is fully gapped with a magnitude of about 2.35 meV. In combination with the theoretical calculations, we attribute the *d*-wave gap to the $Ni\text{-}3d_{x^2-y^2}$ orbital, while the *s*-wave gap to the $Nd\text{-}5d_{3z^2-r^2}$ and/or $Nd\text{-}5d_{xy}$ orbitals. Our results reveal that the NSNO superconductor provides a close analog of the cuprate. This observation shines new light for studies of the newly found Ni-based superconductors.

**Added information of recent progress.** We have made further control experiments on samples with improved surface morphologies. After a long time vacuum annealing, some areas of the surface show layer-by-layer structure with terraces, which makes the STS measurements easier. The conclusions remain the same, namely, we have observed spectra with two dominant superconducting gaps, one *d*-wave and one *s*-wave. In some cases we have observed the spectra with mixed contributions of these two superconducting components. For details, please refer to the Control Experiment section of the Supplementary Information.

First three authors Qiangqiang Gu, Yueying Li & Siyuan Wan contribute equally to the work. Correspondence should be addressed to: ynie@nju.edu.cn & hhwen@nju.edu.cn

**Acknowledgments:** We are grateful to Wei Ku for useful discussions. This work was supported by the National Key R&D Program of China (Grant nos. 2016YFA0300401 and 2016YFA0401704) and National Natural Science Foundation of China (Grants: A0402/13001167, A0402/11774153, A0402/11534005, 1861161004 and A0402/11674164). Y.F.N. also acknowledges the Fundamental Research Funds for the Central Universities (Grant No. 0213-14380167).

**Author contributions:** The NSNO thin films were grown by Y.L., W.G. and Y.F.N., and post-annealed by Q.L., X.Z. The STM measurements and data analysis were done by Q.Q.G., S.Y.W., H.Z.L., H.Y., and H.H.W. The manuscript was written by H.H.W., which is supplemented by others. H.H.W. has coordinated the whole work.

**Competing interests:** The Authors declare no competing interests.

**Figures and legends**

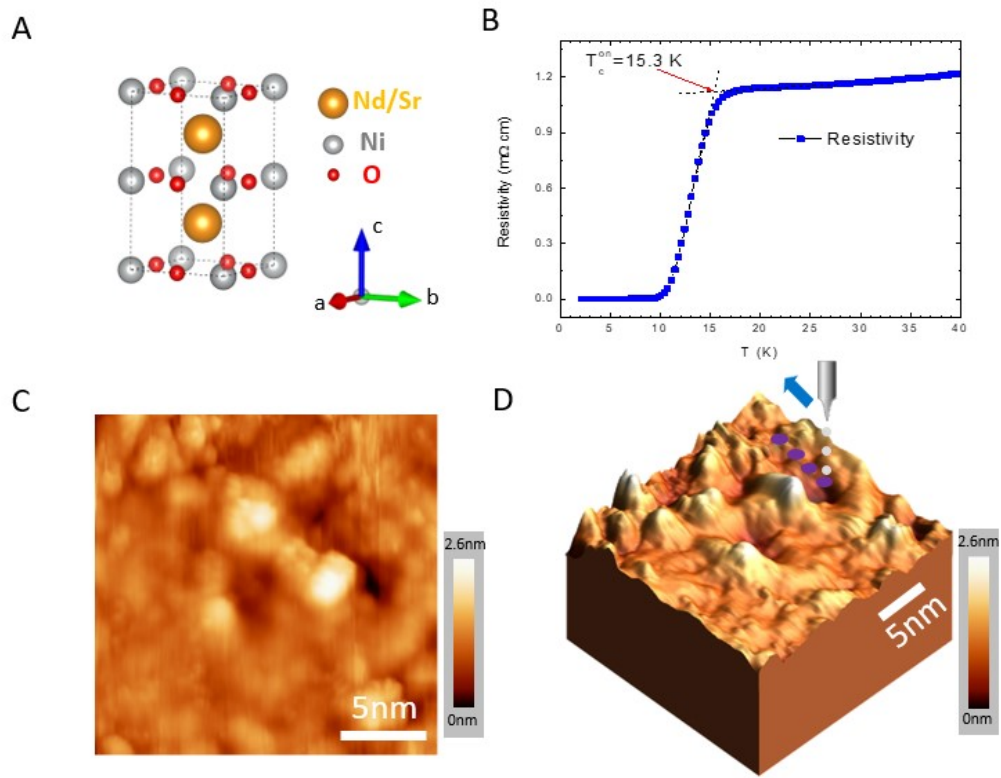

**Figure 1** Resistivity and topographic image of NSNO films. **A,** Schematic plot of the atomic structure of $Nd_{1-x}Sr_xNiO_2$. **B**, Temperature dependence of resistivity measured at zero magnetic field. The onset transition temperature is about 15.3K, and zero resistivity is achieved at about 9.1K. **C, D,** 2D and 3D illustrations of the topographic images measured in an area. One can see that the surface roughness is about 1~2 nm. On the upper-right corner, there are four spots at which we observe the spectra with a V-shape gap. Set conditions for the measurements: *I*= 20 pA, *V*= -5.5 V.

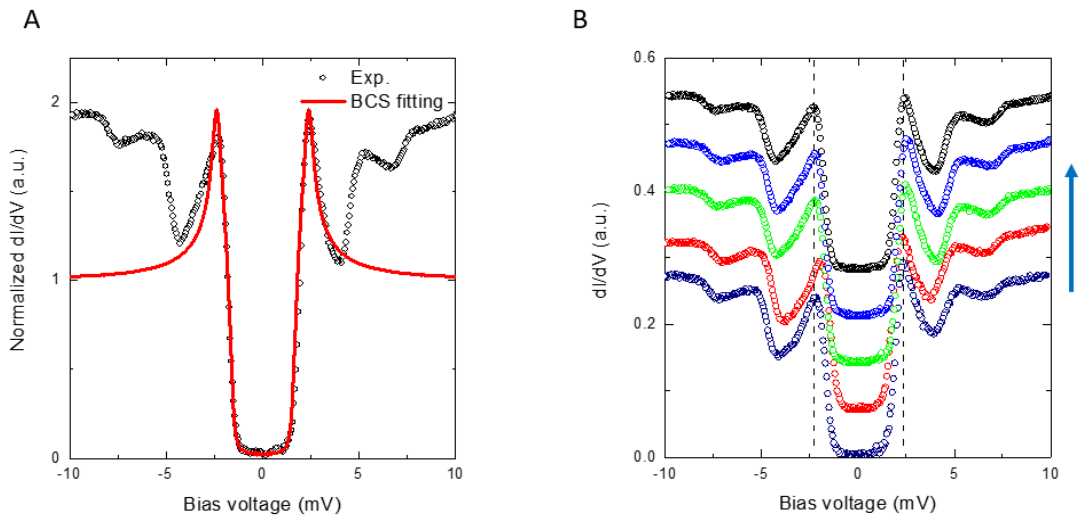

**Figure 2** Tunneling spectra with a full gap. **A,** A tunneling spectra with a full gap (circles) measured at 0.35 K and the Dynes model fit (red solid line). The fitting parameter are as follows, gap function $\Delta_s$ = 2.35(0.15cos4$\theta$ +0.85) meV, scattering rate $\Gamma$ is 0.05 meV, the thermal broadening temperature is 1 K. One can see that the data can be well fitted with the Dynes model. Outside the coherence peaks, there are two side peaks at energies of about 5~6 mV. These might correspond to some bosonic modes. **B,** Spectra measured at different locations in a small area. All show the similar behavior. Setting conditions: I = 100 pA, V = 5 mV.

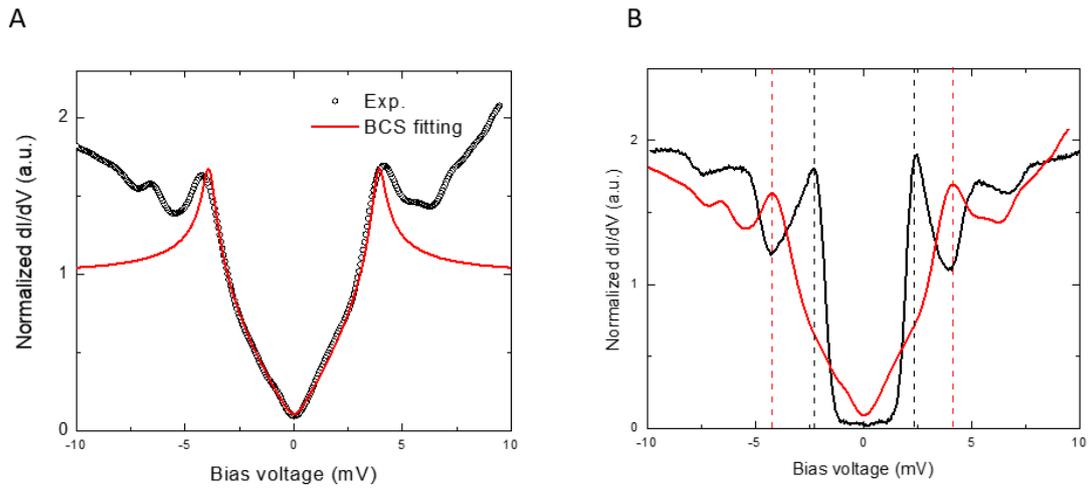

**Figure 3** Tunneling spectra with a V-shape gap and comparison of two kinds of gaps. **A,** A typical tunneling spectrum with a V-shape gap (circles) measured at 0.35 K and the Dynes model fit (red solid line). The fitting is done with a *d*-wave gap with gap function $\Delta_d$ = 3.9cos2$\theta$ meV, scattering rate $\Gamma$ is 0.12 meV, the thermal broadening temperature is 1 K. One can see that the data can be well fitted with the Dynes model. Outside the coherence peaks, there are two small side peaks at energies of about 7 mV. These might correspond to some bosonic modes. **B,** Comparison of the two different spectra with distinct gap structures. On the spectrum with V-shape, double kinks appear at energies of about ±2.35 mV. This strongly indicates that the two gaps are derived from different bands. Measurement setting conditions: I=100 pA, V=5 mV.

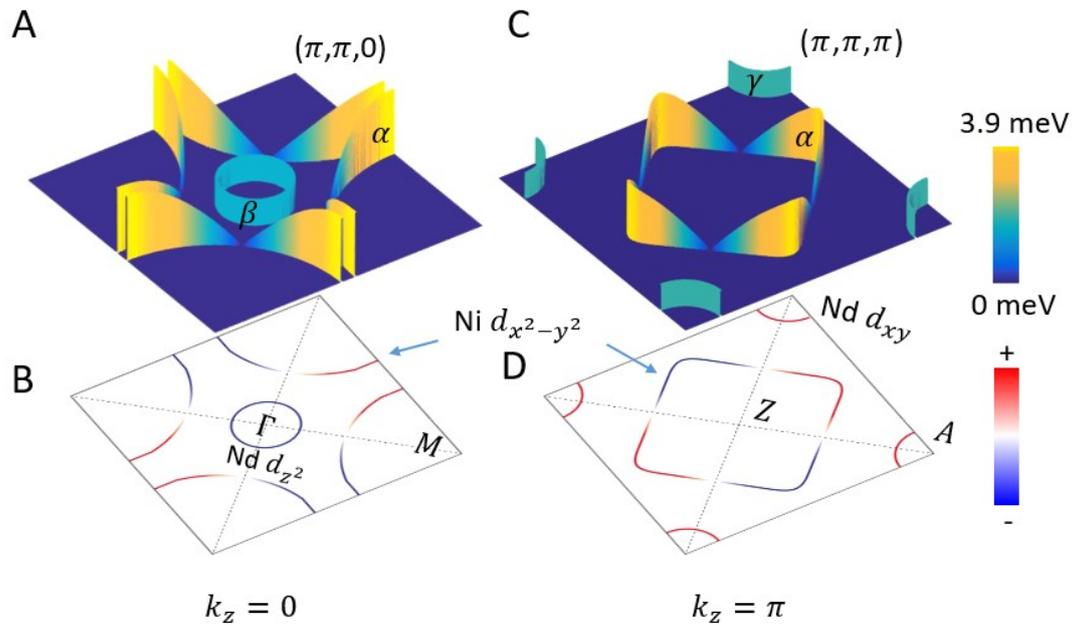

**Figure 4** Cartoon picture for the Fermi surfaces and the gap structure. **A, B,** The Fermi surfaces and the gap amplitude at cut $k_z = 0$. A *d*-wave gap is formed on the $\alpha$ Fermi pocket centered around $\Gamma$, and an *s*-wave gap on the $\beta$ pocket. Actually the Fermi surface on the $\alpha$ pocket at this cut looks very similar to that of underdoped cuprates. The height of the colored walls in **A** represents gap magnitude on each Fermi surface. **C, D,** The Fermi surfaces and the gap amplitude at the cut $k_z = \pi$. A *d*-wave gap is formed on the $\alpha$ Fermi pocket centered around Z, now the Fermi surface becomes a closed square like mimicking that of overdoped cuprate, and an *s*-wave gap on the $\gamma$ pocket around A. The height of the colored walls in **C** indicates the gap magnitude on each Fermi surface. The blue and red colors in **B** and **D** represent the gap signs. Note, about the *s*-wave gaps on the $\beta$ and $\gamma$ pockets, we depict them with different colors.

# Supplementary Information for
# Two superconducting components with different symmetries in $Nd_{1-x}Sr_xNiO_2$ films

Qiangqiang Gu[1], Yueying Li[2], Siyuan Wan[1], Huazhou Li[1], Wei Guo[2], Huan Yang[1], Qing Li[1], Xiyu Zhu[1], Xiaoqing Pan[3], Yuefeng Nie[2]* & Hai-Hu Wen[1]*

## Note 1. Experiment details

The $Nd_{1-x}Sr_xNiO_3$ thin films are grown with reactive molecular beam epitaxy (MBE) technique on $SrTiO_3$ substrates. Then the films are put in an evacuated quartz tube together with a pellet of $CaH_2$, which is followed by a heat treatment at around 340°C for 100 min. The $Nd_{1-x}Sr_xNiO_2$ phase is achieved, which is evidenced by the XRD data and observation of superconductivity after post annealing. For some films we take further annealing at 180°C for 12 hours. After that some areas of the surface show the layer-by-layer structure with terraces.

To check the quality of the $Nd_{1-x}Sr_xNiO_2$ thin film, we have measured temperature dependence of resistivity by a physical property measurement system (PPMS-9T, Quantum Design) with the standard four-probe method. The onset superconducting transition temperature in this sample is about 15.3 K and the zero resistance appears at about 9.1 K.

For post-annealing, a film with dimensions of 5×5 mm² is cut into several pieces, some for the resistive measurements, and others for the STM/STS measurements. For STM/STS measurements, the films are amounted on the special sample holders of STM and quickly transferred to the load-lock chamber for evacuation to avoid possible degradation. STM/STS measurements are carried out in a scanning tunneling microscope (USM-1300, Unisoku Co., Ltd.) with the ultrahigh vacuum up to 10$^{-10}$ torr, low-temperature as 350 mK, and magnetic field up to 11 T. The electrochemically etched tungsten tips are used during all the STM measurements. To raise signal-to-noise ratio in dI/dV measurements, the set points for spectra acquired within ±10 mV are 5 mV and 100 pA, meanwhile a typical lock-in technique is used with an ac modulation of 0.1 mV and 987.5 Hz. All data are taken at either 1.5 K or 0.35 K.

**Note 2. Dynes model fitting**

Based on the Dynes model (*1*), the measured tunneling current between a metallic tip (providing constant density of states near Fermi energy) and an anisotropic superconductors can be expressed as

$$I(V) = \frac{1}{2\pi}\int_0^{2\pi} d\theta \int_{-\infty}^{+\infty} d\varepsilon [f(\varepsilon) - f(\varepsilon + eV)] \cdot \text{Re}\left(\frac{\varepsilon + eV + i\Gamma}{\sqrt{(\varepsilon + eV + i\Gamma)^2 - \Delta^2(\theta)}}\right).$$

(S1)

Here *f*(ε) is the Fermi distribution function containing the information of temperature. We set the practical electronic temperature as 1 K which is a little higher than the nominal one displayed by thermometer. The scattering factor $\Gamma$ denotes the inverse quasiparticle lifetime in unit of meV and $\theta$ represents the azimuth angle along the Fermi surface in the Brillouin zone. The best fittings to the data shown in Fig.2 and 3 yield the fitting parameters which are given in Supplementary Table 1.

**Supplementary Table 1. Dynes model fitting parameters**

| Gap notation | Gap function | $\Gamma$ (meV) | Thermal braodenning temperature |
|---|---|---|---|
| *s*-wave | $\Delta_s = 2.35(0.15\cos4\theta + 0.85)$ meV | 0.05 | 1 |
| *d*-wave | $\Delta_d = 3.9 \cos2\theta$ meV | 0.12 | 1 |

**Note 3. Control experiments on other samples with different morphologies**

We have successfully repeated the primary results presented in Fig.2 and Fig.3 in the main text on other $Nd_{1-x}Sr_xNiO_2$ thin films with different morphologies. The smaller s-wave gap and the larger d-wave gap can be found at different locations on these thin films. Actually, after the topotactic reduction with $CaH_2$, the atomically flat surface of the original MBE deposed thin film could be damaged significantly, leading to the roughness of about 1~2 nm, as shown in Fig.1**D**. In this case, it is not easy to stabilize the tunneling junction between the tip and the sample. Thus, we need more flat surfaces for further measurements. For that purpose, we anneal the thin films at about 180℃ in ultrahigh vacuum ($10^{-9}$ torr) for 12 hours. As shown in Supplementary Fig.4 and Fig.5, we find a much more flat surfaces with layer-by-layer structure, in contrast to the topography shown in Fig.1**D** of the main text.

During the control experiments on measuring the spectra, we find that the probability of measuring an s-wave gap is larger than the *d*-wave gap before the long time vacuum annealing. The situation seems to be different after this new annealing. We can easily find the spectra with *d*-wave gap on the flat surface as shown in Supplementary Figure 6A. Sometimes, as shown in Supplementary Figure 6B, we can also see the spectra with mixed contributions of the two-gap component, featuring a large *d*-wave gap with an smaller *s*-wave gap. Supplementary Fig. 7A and 7B show the single *d*-wave and single *s*-wave spectra measured at different locations.

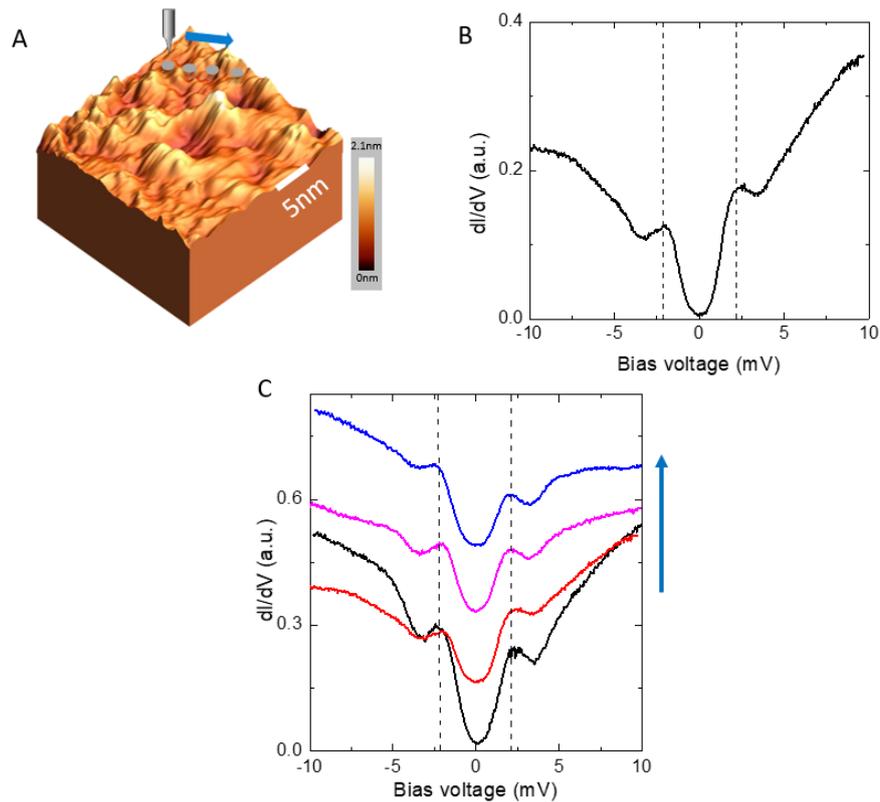

**Supplementary Figure 1** Topographic image and tunneling spectra featuring the smaller gap acquired at 1.5 K. **A**, 3D view of topographic image of $Nd_{1-x}Sr_xNiO_2$ thin film based on the STM data. At the four spots marked here we measure the spectra shown in **C**. **B**, Tunneling spectrum measured at the second marked spot in **A**. The vertical dashed lines indicate the coherence peak positions. **C**, A series of spectra measured on the marked spots. We visualize a nearly full superconducting gap with the coherence peaks located at about ±2 mV. The spectra are offset for clarity. Set-point conditions for **A** are $V_{set}$ = -5.5 V, $I_{set}$ = 20 pA; and for **b,c** are $V_{set}$ = 5 mV, $I_{set}$ =100 pA.

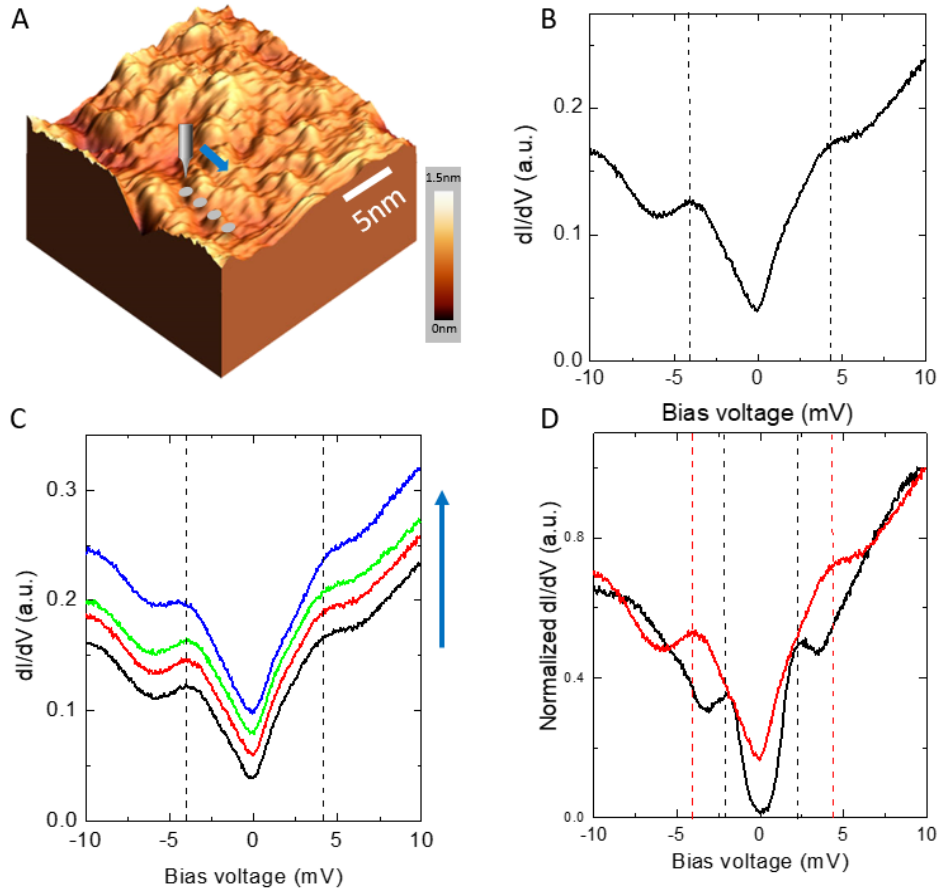

**Supplementary Figure 2** Topographic image and tunneling spectra featuring the larger gap acquired at 1.5 K. **A**, 3D view of topographic image of Nd$_{1-x}$Sr$_x$NiO$_2$ thin film based on the STM data. **B**, Tunneling spectrum measured at the second marked point in **A**. **C**, A series of spectra measured at the marked points in **A**. We see a clear suppression of the low-energy spectral weight and a pair of kinks at about ±3.9 mV which should correspond to the coherence peaks. The coherence peaks are strong suppressed because of the pairing-breaking scattering of disorders and defects, as well as the thermal broadening effect. A finite residual spectral weight near zero energy indicates the quasi-particle excitations from nodal region of a *d*-wave gap. The spectra are offset for clarity. **D**, Comparison between the spectra with small gap and large gap measured at 1.5 K. Set-point conditions for **A** are $V_{set}$ = 7.2 V, $I_{set}$ = 50 pA; and for **B**,**C** are $V_{set}$ = 5 mV, $I_{set}$ = 100 pA.

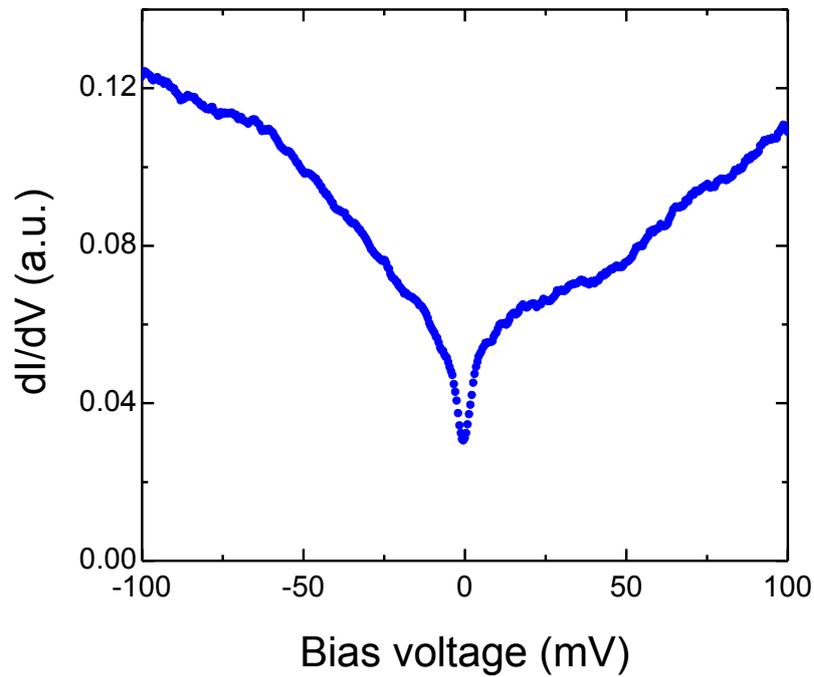

**Supplementary Figure 3** Tunneling spectra acquired within ±100 mV at 1.5 K. A V-shaped background indicates a bad metal behavior in the normal state of $Nd_{1-x}Sr_xNiO_2$. The suppression of spectral weight near zero energy is induced by the formation of superconducting gaps. The seemingly weak signature of the superconducting gap on this spectrum is due to two reasons. One is that we change the set-point conditions as $V_{set}$ = 100 mV, $I_{set}$ = 100 pA which sets a relatively longer distance between the tip and sample, experiencing an exponential decay of superconducting order parameter at the end of the tip. Another is that we increase the AC oscillation amplitude to enhance the dI/dV signal while it inevitably smears out the low-energy details of the spectrum.

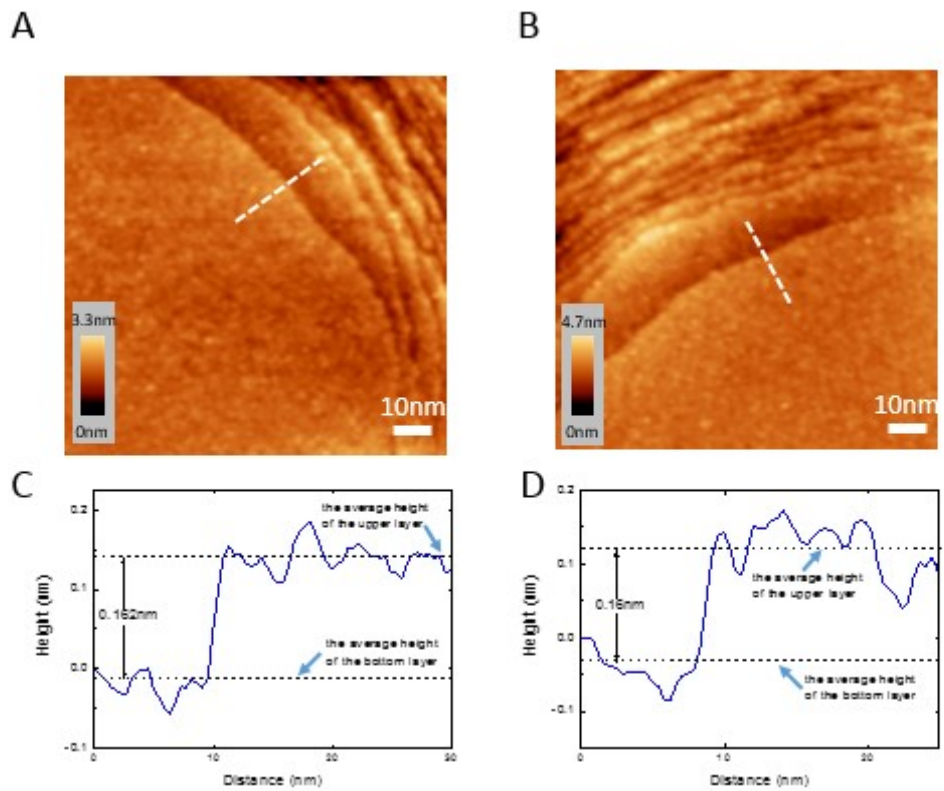

**Supplementary Figure 4** Topographic image after taking further annealing of the sample at about 180°C in ultrahigh vacuum for 12 h. **A**, Topographic image of $Nd_{1-x}Sr_xNiO_2$ thin film in an area of 100 nm×100 nm. We can see a much more flat surface, with actually a layer-by-layer structure with terraces. **B**, Topographic image of another area of 100 nm×100 nm. **C,D** Spatial distribution of the height measured along the dashed line in **A** and **B**. The step height is about 0.16 nm, being consistent with one half of the unit cell height. Taking the structure factors into account, we believe the top surfaces exposed may correspond to either Nd/Sr layer or $NiO_2$ layer at different locations. Set-point conditions for **A** and **B** are $V_{set}$ = 6.3 V, $I_{set}$ = 50 pA.

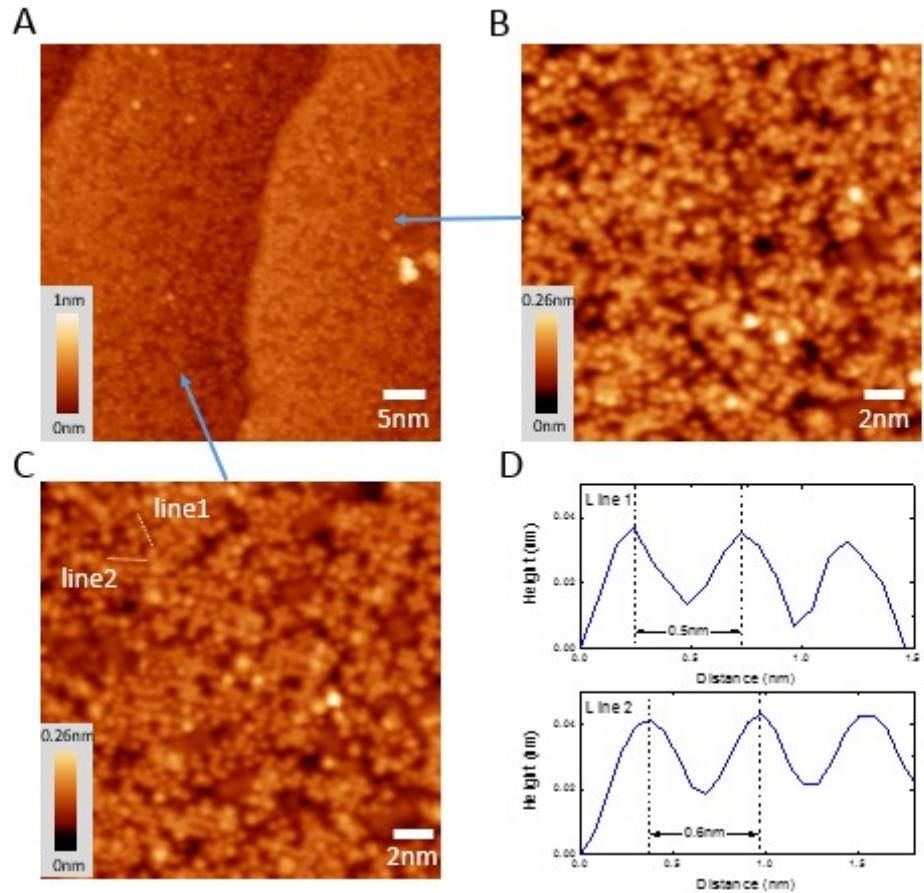

**Supplementary Figure 5** Atomically flat topographic image after the long time vacuum annealing process. **A**, Topographic image of $Nd_{1-x}Sr_xNiO_2$ thin film in an area of 40 nm×40 nm. We can see a clear step with the height of about half unit cell. **B**, Topographic image of the upper layer. **C**, Topographic image of the bottom layer. **D**, Spatial distribution of the height measured along the lines crossing several white spots in **C**. Based on the atomic structure of $Nd_{1-x}Sr_xNiO_2$, we understand that there are no natural and neutral cleaving planes, thus the topo layer may be constructed by Nd/Sr and $NiO_2$ planes at different locations. On the surface we see many white spots, we think they are the Nd/Sr atoms. As shown in **D**, the distances between the adjacent white spots are between 0.5 nm to 0.6nm, which are much larger than the in-plane Ni-Ni or Nd-Nd lattice constant 0.39 nm, but rather close to the $\sqrt{2}\times\sqrt{2}$ structure of them. Thus we think locally these Nd/Sr atoms may try to reconstruct into this structure. Set-point conditions for **A**, **B** and **C** are $V_{set}$ = 3.6 V, $I_{set}$ = 50 pA.

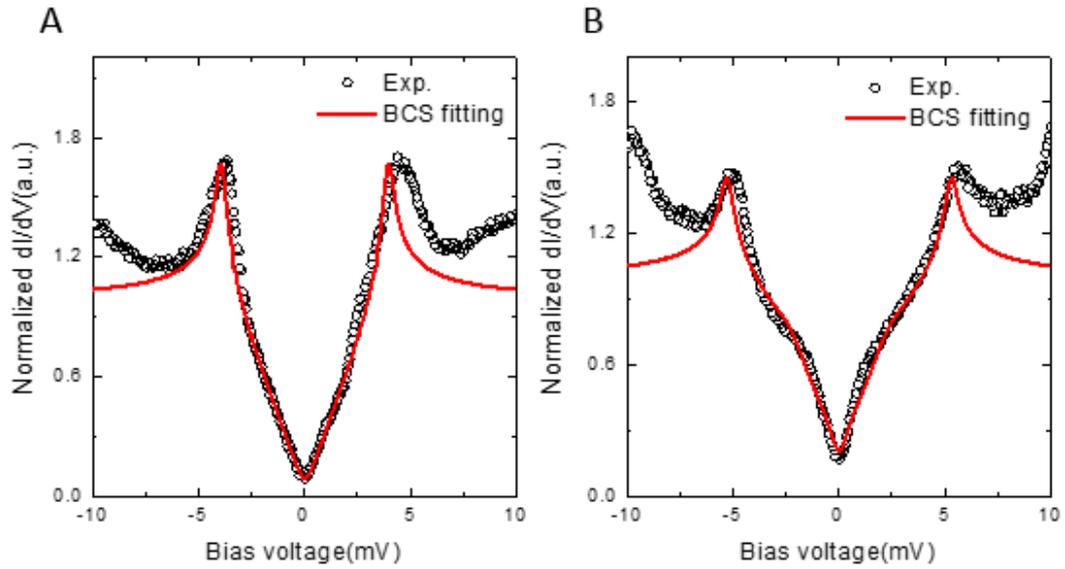

**Supplementary Figure 6** The larger *d*-wave superconducting gap measured on Nd$_{1-x}$Sr$_x$NiO$_2$ thin film at 0.4 K after long time vacuum annealing. **A**, A typical spectrum with *d*-wave gap measured on the flat surface shown in **Supplementary Figure 5B**. The fitting is done with a *d*-wave gap with gap function $\Delta_d$ = 3.95(0.95cos2$\theta$+0.05cos6$\theta$) meV, the scattering rate $\Gamma$ is 0.07 meV, the thermal broadening temperature is 1 K. **B**, A typical spectrum with mixed contribution of two gaps. The fitting parameters are $\Delta_1$= 5.3*[0.8cos(2$\theta$)+0.2cos(6$\theta$)] (meV), $\Gamma_1 = 0.1$, $\Delta_2$= 2 meV, $\Gamma_2 = 0.7$ with p$_1$ = 85% and T = 1 K. Experimental data are denoted by black circles and the Dynes model fitting results are denoted by red curves. Set-point conditions for **A**, **B** are $V_{set}$ = 4 mV, $I_{set}$ = 100 pA.

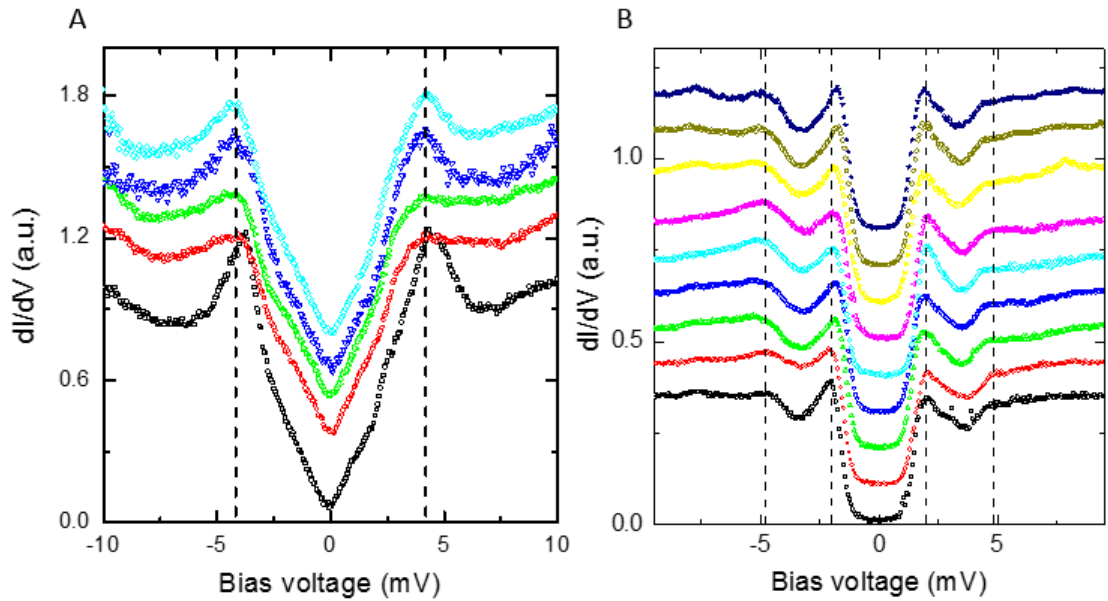

**Supplementary Figure 7** Control experiment of spectra with *d*-wave and *s*-wave superconducting gaps measured at 0.4 K. **A**, A series of spectra with *d*-wave gap with gap maxima of about 4.2 meV. **B**, A series of spectra with *s*-wave gaps with magnitude of about 2 mV. These two types of spectra are measured at different locations on the thin film. Set-point conditions is $V_{set}$ = 4 mV, $I_{set}$ = 100 pA for **A** and $V_{set}$ = 3 mV, $I_{set}$ = 100 pA for **B**.